\documentclass[preprint,review,12pt]{elsarticle}

\usepackage{fontenc}

\usepackage{subfigure}
\usepackage{amsmath}
\usepackage{subeqnarray}
\usepackage{tikz} 
\usepackage{ctable}

\usepackage{dcolumn}
\usepackage{array}
\usepackage{amssymb}
\graphicspath{{./img/}}
\usepackage{graphicx}

\usepackage{natbib}
\usepackage{color}
\usepackage{amsmath}
\usepackage{fancyhdr}
\usepackage{textcomp}

\usepackage{marvosym}
\usepackage{nomencl}

\usepackage{epstopdf}
\usepackage{hyperref}

\journal{Proceedings of the 35$^{th}$ Combustion Symposium}

\newcommand{\mratio}{\left(\dfrac{m^+}{m^-}\right)^{1/2}}

\usepackage{graphicx,color,psfrag}

\newcommand{\p}{\partial}
\newcommand{\rl}{\rightarrow}

\biboptions{comma,square,sort&compress}

\makenomenclature

\usepackage{ifthen}
\renewcommand{\nomgroup}[1]%
	{\ifthenelse{\equal{#1}{S}}{{\item[\textbf{Subscripts}]}{}}}

\begin{document}
\biboptions{comma,square,sort&compress}
\nomenclature[e]{$\alpha $}{relaxation parameter}
\nomenclature[d]{$Y' \: [Y]$}{mass fraction [dimensionless]}
\nomenclature{$S_{L}$}{Flame Speed}
\nomenclature{$\rho$}{mixture density}
\nomenclature{$c_{p}$}{heat capacity}
\nomenclature{$D$}{diffusivity}
\nomenclature[d]{$E' \: [E]$}{electric field [dimensionless]}
\nomenclature{$e$}{elementary charge}
\nomenclature{$E_{a}$}{activation energy}
\nomenclature{$c$}{concentration}
\nomenclature{$m$}{mass of particle}
\nomenclature[e]{$\gamma$}{dimensionless heat release}
\nomenclature{$V$}{diffusion velocity}
\nomenclature{$Z$}{charge}
\nomenclature[e]{$k$}{iteration counter}
\nomenclature{$\nu$}{mobility}
\nomenclature[d]{$T [ \theta ]$}{temperature  [dimensionless]}
\nomenclature{$A$}{frequency factor}
\nomenclature[d]{$\Omega \: [\omega]$}{reaction rate [dimensionless]}
\nomenclature[d]{$x' \: [x] $}{spatial coordinate [dimensionless]}
\nomenclature[e]{$\mu$}{eigenvalue of dimensionless conservation equations}
\nomenclature[d]{$q \: [Q]$}{heat of reaction [dimensionless]}
\nomenclature[e]{$\beta$}{Zeldovich number}
\nomenclature[e]{$Le$}{Lewis number}
\nomenclature[e]{$\cal{A}$}{dimensionless frequency factor for reaction III}
\nomenclature[e]{$\cal{B}$}{dimensionless frequency factor for reaction IV}
\nomenclature[e]{$\Delta \beta$}{dimensionless activation energy increment}
\nomenclature[d]{$\varepsilon_{0} \: [ \varepsilon ] $}{vacuum permitivity [dimensionless]}
\nomenclature{$W$}{molecular weight}
\nomenclature{$R_g$}{gas constant}
\nomenclature[S]{$Z$}{neutral radical}
\nomenclature[S]{$F$}{fuel}
\nomenclature[S]{$Z^+$}{positive radical}
\nomenclature[S]{$e^-$}{electron}
\nomenclature[S]{$0$}{initial or imposed value}
\nomenclature[S]{$T$}{temperature, thermal}
\nomenclature[S]{$\infty$}{final value}

\begin{titlepage}
\begin{center}

\textsc{\LARGE Effect of an External Electric Field on the Propagation Velocity of Premixed Flames}\\[.5cm]


{\Large Mario S\'anchez--Sanz\textsuperscript{a}, Daniel C. Murphy\textsuperscript{b}, C. Fernandez-Pello\textsuperscript{b}}


{\itshape \textsuperscript{a} Dept. Ingenier\'ia T\'ermica y de Fluidos,
Universidad  Carlos III de Madrid}


{\itshape \textsuperscript{b}Department of Mechanical Engineering, University of California Berkeley}

\vspace*{.25\baselineskip}


\vspace*{.25\baselineskip}

{\textbf{Corresponding Author} Daniel C. Murphy\textsuperscript{b}}

{60 Hesse Hall, Department of Mechanical Engineering, University of California Berkeley}


{Tel: (415) 613-6270}


{Fax: (510) 642-1850}


{email: dmurphy0516@gmail.com}

\vspace*{.15\baselineskip}

{\textbf{Word Count:} (determined via Method 2 - LaTeX formatting) \\Method 2: Created a complete two-column formatted version of the paper\\ \textbf{Pages: 6, 1} partial page with single column \textbf{205 mm} in length\\ \textbf{Total: 5850 words}}

\vspace*{.15\baselineskip}

{\textbf{Preferred colloquium topic area:} 12. NEW TECHNOLOGY CONCEPTS, REACTING FLOWS AND FUEL TECHNOLOGY, 4. LAMINAR FLAMES }

\vspace*{.15\baselineskip}

{\textbf{Color reproduction:} Not Required}

\end{center}
\end{titlepage}

\begin{frontmatter}

\title{ Effect of an External Electric Field on the Propagation Velocity of Premixed Flames}


\author[auth1]{M. S\'anchez--Sanz\corref{cor1}} \ead{mssanz@ing.uc3m.es}
\author[auth2]{D. Murphy \corref{cor2}} \ead{dmurphy0516@gmail.com}
\author[auth2]{C. Fernandez-Pello \corref{cor3}} \ead{ferpello@me.berkeley.edu}

\address[auth1]{Dept. Ingenier\'ia T\'ermica y de Fluidos,
Universidad  Carlos III de Madrid, 28911, Legan\'es, Spain}

\address[auth2]{Department of Mechanical Engineering The University of California Berkeley}

\cortext[cor2]{
Corresponding author. 
Tel: (415) 613-6270}
\noindent

\begin{abstract}
There have been many experimental investigations into the ability of electric fields to enhance combustion by acting upon ion species present in flames \cite{Murphy2013}.  In this work, we examine this phenomenon using a one-dimensional model of a lean premixed flame under the influence of a longitudinal electric field.  We expand upon prior two-step chain-branching reaction laminar models with reactions to model the creation and consumption of both a positively-charged radical species and free electrons. Also included are the electromotive force in the conservation equation for ion species and the electrostatic form of the Maxwell equations in order to resolve ion transport by externally applied and internally induced electric fields. The numerical solution of these equations allows us to compute changes in flame speed due to electric fields.  Further, the variation of key kinetic and transport parameters modifies the electrical sensitivity of the flame.  From changes in flame speed and reactant profiles we are able to gain novel, valuable insight into how and why combustion can be controlled by electric fields.
\end{abstract}


\end{frontmatter}


\printnomenclature
\section{Introduction}

In order to study the effect of the electric field on a premixed flame, this
paper will consider a one-dimensional model of a lean premixed flame under the
influence of a longitudinal electric field. The electric field is oriented in the direction of the
gas flow and can have positive or negative signs, indicating different electric field polarities. 
The model is defined by the set of conservation equations and chemical reactions which should
reflect the behavior of a lean premixed flame subjected to an electric field. To
model the interaction between the flame and the electric field, we need to include in the chemistry
model a radical that can be ionized at high temperatures to give a protons and electrons that can
be affected by the electric field. 

A good candidate is the the two-step, chain-branching chemistry model developed originally by \cite{Zeldovich1948,Zeldovich1985} and further developed later by \cite{Linan1971} using the high activation energy asymptotic. The model was later slightly modified in \cite{Dold2002,Dold2007} by linearizing the radical recombination step,
allowing the definition of an explicit  crossover temperature below which the chain-branching
reaction remains frozen. Some authors refer this two-step kinetic model
as a better approach to real hydrocarbons and hydrogen flames description
than the one-step model \cite{Dold2002,Dold2007, Sharpe2009}. Chain-branching reactions are
typically located in the high-temperature region of the flame because of its high activation energy.
These reactions produce an increase of the intermediate
species or chemical radicals which diffuse upstream and downstream of this
thin chain-branching zone and recombine by means of a very exothermic
chain-termination reaction in a wider region. Therefore, the heat release occurs throughout the
flame and fuel exhaustion is reached interior to the flame
\cite{Dold2007, Sharpe2009}, a real characteristic of flames. In the one-step model fuel
consumption and final adiabatic flame is reached all at once. \\
To complement the model proposed by Dold, we incorporate here two additional reactions to account
for the production and consumption of charged species. According to \cite{Pedersen1993}, the source
of ions is generally accepted to be the chemiionization reaction in which a radical reacts with a
third body to give a proton and an electron. Later, the proton would recombine with the
electrons via a dissociative recombination to release a certain amount of heat. When an external
electric field is imposed across the flame front, the charged particles are removed from the
reaction zone at a rate proportional to the electric field strength until the field strength is
large enough to make the removal rate of electrons and protons equal to the chemiionization
formation rate, leading to current saturation.\\
\autoref{table:reactions} shows the reactions used to model the effect of the electric field in a
premixed flame. As we anticipated above, reactions I and II replicate the
model proposed by Dold while reaction III and IV model the chemiionization and the dissociative
recombination, respectively, first proposed by \cite{Calcote1960,Calcote1965} and used in multiple numerical studies \cite{Pedersen1993}, \cite{Papac2007} and \cite{Belhi2010}.  

\label{sec:Intro}

\section{Formulation}

Consider a planar premixed flame propagating with a constant velocity, $S_L$, with respect to an
unburned gas at initial temperature $T_0$ and fuel mass fraction $Y_{F_0}$. The mixture is
assumed to be deficient in fuel and the mass fraction of the oxidizer, which is in
abundance, remains nearly constant. For the sake of simplicity, the paper deals with a
diffusive-thermal model, according to which $\rho$, $c_p$, $D_T$, $D_F$ , $D_Z$
are all constant.\\
The equations describing the structure of this flame in
the presence of an electric field include the mass and species conservation equations. Additionally,
the model needs to consider the effect that the electric field exerts on the charged species. Since
the concentrations of charged species are usually small, we will assume that the relation defining
the flux of the {\textit i}th species due to diffusion plus electromigration will be that for a
weakly-ionized plasma, an extreme that considerably simplifies our treatment
of the problem and which is usually the case in real flames \cite{Cancian2013}. Thus, the
contributions to the flux of charged species can be linearly superposed to define the Nerst-Planck
equation describing the interaction between the electric field and the charged species
\cite{Probstein1995}. \\
With currents and electric fields present, the laws of electrodynamics should be incorporated, in
the form of the Maxwell equations, into the equations of mass and energy conservation. Nevertheless,
if there are no magnetic fields  and the electric field does not change with time, the electrodynamic
problem reduces to an
electrostatic one. In this case, the electric field can be computed as $\p E' /\p x= (c_{+} -
c_{-}) e/ \varepsilon_0 $. \\
In an electrically neutral ionized gas, the condition $\sum c_i Z_i \simeq 0$ is satisfied, where
$Z_i$ and $c_i$ are  the charge and the concentration of the particle
$\textit{i}$, respectively. Since the electrons represent
nearly 90 \% of the negative charge carriers \cite{Fialkov1997}, we can write the
electrons-to-protons characteristic mass fraction ratio as $Y_{e^-,0}/ Y_{Z^+,0} \simeq
m^-/m^+ \ll 1$, indicating that the contribution of the electrons to the total mass of the gas is
small. \\
In the constant density approximation we are considering here, the problem reduces to the
integration
of energy and mass transport equations for neutral and charged species. Unlike neutral gases,
where diffusion is controlled by Fick's law, the presence of an electric field can change the way in
which the particles are redistributed in an inhomogeneous mixture. Even when no external
electric field is applied, a displacement of charged particles would create a charge imbalance
that, in turn, would induce an electric field opposing the charge displacement.
This effect can be taken into account by defining the diffusion velocity $V^k$ as
\cite{Pedersen1993,Probstein1995}
\begin{equation}
 \rho Y V^k_j = - \rho D_j \dfrac{\p Y_i}{\p x} + Z_k \rho \nu^k Y_i E,
 \label{eq:drift}
\end{equation}
where $Z_k$ is negative if the species is negatively charged, positive if the species is
positively charged and zero if the species is neutral. The diffusion coefficient of the neutral
species is considered constant, being $D_F$ and $D_Z$ the diffusion coefficients of fuel and
radical Z, respectively. On the other hand, Belhi et al  \cite{Belhi2010} introduced, following
Delcroix \cite{Delcroix1963}, the following expression for the ratio between the diffusion
coefficient of electrons and ions 
$D_{e^-} = D_{Z^+} (m^+/m^-)^{1/2}$
where $m^+$ and $m^-$ are the mass of a single proton and
electron respectively. \\
The mobility of a charged particles $\nu_i$ is defined as the ratio between its drift velocity and
the electric field strength.  The strict calculation of the mobility of ions $\nu_{Z^+}$ and
electrons $\nu_{e^-}$ would imply the appropriate assessment of the effect of the temperature
and concentration changes  on $\nu_i$ \cite{Papac2007}. Nevertheless, and for the sake of
simplicity, we will assume hereafter constant proton and electron mobilities. Furthermore, we can
write the ratio between the mobilities of electrons and ions by using the Einstein relationship
given in \cite{Belhi2010}
$\nu_{e^-} = \nu_{Z^+}(m^+/m^-)^{1/2}$\\
The four-step, chain-branching kinetic mechanism used here to model the effect of the electric
field includes the autocatalytic and recombination steps given above in Table
\ref{table:reactions}, where $\Omega_I$ is the temperature-sensitive, chain-branching reaction
rate, with $E_I$ the activation energy and $A_I$ the frequency factor, and $\Omega_{II}$ is the
temperature-independent completion reaction rate, with $A_{II}$ the rate constant. Notice that, as
indicated by \cite{Dold2007}, the reaction can only take place if the temperature $T>T_c$, where
$T_c$ is the branching temperature that takes into account the amount of radical removed
by diffusion from the inner branching zone and is obtained by imposing $\Omega_I=\beta^2
\Omega_{II}$ to give 
\begin{equation}
\dfrac{A_I}{A_{II}} \dfrac{W}{W_F} Y_{F_0} = \left\{\dfrac{E}{R_g} \dfrac{T_c-T_0}{T_c} \right\}^2
e^{E_I/R_g T_c}
\end{equation}
This temperature is used here to define the non-dimensional temperature $\theta=(T-T_0)/(T_c-T_0)$
and the Zel'dovich number $\beta= (T_c-T_0)/(R_g T_0^2)$. \\ 
A great amount of work has been done to identify the ions species present in a flame and the
mechanisms responsible for their production. A detailed account can be found in
\cite{Calcote1960,Calcote1965,Pedersen1993}. In this regard we incorporate in our simplified model
the steps III and IV that account for the temperature-dependent chemiionization of the radical Z and
the posterior exothermic recombination of protons and electrons.\\
As a summary of the exposed above, we introduce in the energy and mass conservation equations the
non-dimensional temperature $\theta$ and spatial coordinate  $x=x'/(D_T/S_L)$ and the
scaled mass fractions of fuel  $Y_F=Y_F'/Y_{F_0}$,radical $Y_Z=Y_Z'/(W_Z (Y_{F_0}/W_F)$, protons
$Y_{Z^+}=Y_{Z^+}'/(Y_{F_0} \left(W_Z /W_F\right))$ and electrons $Y_{e^{-}}=Y_{e^{-}}'/(Y_{F_0}
(m^{-}/m^{+}) \left(W_Z /W_F\right) )$ to yield
the non-dimensional conservation equations
\begin{align}
\dfrac{d \theta}{d x} &  =  \dfrac{d^2 \theta}{d x^2} + \mu Q \left(\omega_{II}  +
\omega_{IV} \frac{q_{IV}}{q_{II}} \right)  \label{eq:theta_eq} \\
\dfrac{d Y_{F}}{d x} & =  \dfrac{1}{Le_F} \dfrac{d^2 Y_F} {d x^2} - \mu \omega_I
\label{eq:F_eq} \\
 \dfrac{d Y_{Z}}{d x} & = \dfrac{1}{Le_Z} \dfrac{d^2 Y_Z} {d x^2}  +  \mu 
\left[\omega_I-\omega_{II} - \omega_{III} \right]
\label{eq:Z_eq} \\
\dfrac{d Y_{Z^{+}}}{d x}  & =-\mu^{1/2} \dfrac{d \left(E Y_{Z^{+}} \right)}{d x}+
\dfrac{1}{Le_{Z^{+}}} \dfrac{d^2 Y_{Z^{+}}}{d x^2} 
 +\mu \left[ \omega_{III}-\omega_{IV} \right] \label{eq:proton_eq} \\
\dfrac{d Y_{e^{-}}}{d x}  & =  \mu^{1/2}  \mratio \dfrac{d (E Y_{e^{-}})}{d x}+ \nonumber
\\
&\qquad+\left(\frac{m^{+}}{m^{-}} \right)^{1/2} \dfrac{1}{Le_{Z^{+}}} \dfrac{d^2 Y_{e^{-}}}{d x^2} 
 +\mu \left[  \omega_{III}-\omega_{IV} \right] \label{eq:electron_eq}
\end{align}
with boundary conditions $\theta=Y_Z=Y_{Z^+}=Y_{e^-}=Y_F-1=0$ at $x \to -\infty$ and
$\theta'=Y'_F=Y'_Z=Y'_{Z^+}=Y'_{e^-}=0$  at $x \to \infty$.\\
The solution of the problem provides the eigenvalue 
\begin{equation}
\mu = \dfrac{\rho A_{II} D_T}{S^{2}_L W }
\end{equation}
 which determines completely the flame velocity $S_L$. Also, the following non-dimensional parameters
appear in the above formulation: the Zel’dovich number $\beta=10$, the dimensionless
heat of reaction $Q = q_{II} Y_{F_0}/[c_p(T_c - T_0) W_F]$, with $q_{II}$ the total heat
released from reactions II, the Lewis numbers of fuel
$Le_F= D_T /D_F$ and radical $Le_Z= D_T / D_Z$ and the heat release parameter $\gamma = (T_c -
T_0)/T_c=0.7$.\\
The non-dimensional reaction rates are written as
\begin{align}
 \omega_I & =\beta^2 Y_Z Y_F   \exp \left\{ \beta \dfrac{\theta-1}{1+\gamma (\theta-1)} \right\}\\
\omega_{II} &= Y_Z \\ 
\omega_{III} &= \beta^2 {\cal{A}} Y_Z  \exp \left\{ (\beta+ \Delta \beta)  \dfrac{\theta-1}{1+\gamma
(\theta-1)}\right\} \\
\omega_{IV} & = {\cal B} Y_{Z^{+}} Y_{e^{-}} 
\end{align}
with ${\cal{A}} = \dfrac{A_{III}}{A_I} \dfrac{W_{Z^{+}} W_F}{W_Z \bar{W}}
\dfrac{e^{-\Delta \beta/\gamma}}{Y_{F_0}}$, ${\cal{B}} =
\dfrac{A_{IV}}{A_{II}}\dfrac{\bar{W} W_Z}{W_{Z^{+}} W_F} Y_{F_0}$ and 
$\Delta \beta= \gamma \dfrac{E_{III}-E_I}{R_g T_c}$ representing the effect of a differential
activation energy between the chain-branching and the chemiionization steps.
In real flames, the heat released through the termination reaction $q_{II}$ is different to that
released through the dissociative recombination  $q_{IV} \ne q_{II}$. Nevertheless, and
for the sake of simplicity, we will assume
hereafter that $q_{IV}=q_{II}$. In this case, the system of equations described above in
(\ref{eq:F_eq})-(\ref{eq:electron_eq}) admits a first integral 
that allows the calculation of the temperature downstream of the reaction region 		
$\theta_\infty=Q (1-F_\infty)$, where $F_\infty$ is the fuel leakage at $x \to \infty$,
facilitating the physical interpretation of the parameter $Q$. Notice that if $q_{IV}/q_{II} \ne 1$
the maximum flame temperature is diminished due to the reduction of the
radical Z that is consumed through reaction II, and the maximum flame temperature will be given by
$\theta_\infty=Q \left[ 1-F_\infty - \mu {\cal B} \left(1-q_{IV}/q_{II} \right) \int_{-
\infty}^\infty Y_{Z^+} Y_{e^-} dx \right] $.  \\
The spatial distribution of non-dimensional electric field \\
$E= E' /[\nu^{+} (\rho D_T A_{II}/\bar{W})^{1/2}]$ depends on the spatial distribution of the charged species and is given by 
\begin{equation}
 \dfrac{d E}{d x}= \dfrac{\mu^{1/2}}{\varepsilon} (Y_{Z^+} - Y_{e^-}) \ \ \ E(x \to
-\infty)=E_0 
\label{eq:E}
\end{equation}
with $\varepsilon= \nu^+ (\varepsilon_0/e) (A_{II} W_{Z^+}) /(\bar{W} Y_{F_0} W_Z/W_F)$ and $E_0$ the
external electric field applied.  \\
The large electrons mobility anticipates an effective diffusion of the electrons away from the
flame. In order to satisfy the boundary conditions specified above, the limits of the
computational domain must reach distances of the order $\mid x \mid \sim (m^+/m^-) \gg 1$.
Nevertheless, an asymptotic approximation at $x \to -\infty$ of
$Y_{e^-}$ and the associated induced electric field $E$ can be derived from eqs.
(\ref{eq:electron_eq}) and (\ref{eq:E}) by imposing $Y_{Z^+}=w_{III}=w_{IV}=0$ to give 
\begin{align}
1-\mu^{1/2} (m^+/m^-)^{1/2} E &= \dfrac{(m^+/m^-)^{1/2}}{Le_{Z^+}}Y_{e^-}^{-1} \dfrac{\p
Y_{e^-}}{\p x} \label{eq:Yeinfty}
\end{align}
with
\begin{multline}
1-\mu^{1/2} (m^+/m^-)^{1/2} E =\\
\sqrt{ \left(1- \dfrac{\mu^{1/2} (m^+/m^-)^{1/2}}{2} E_0 \right)^2 -2 \dfrac{\mu (m^+/m^-)}{\varepsilon Le_{Z^+}} Y_{e^-}} 
\end{multline}
The straightforward integration
of (\ref{eq:Yeinfty}) gives the asymptotic behavior of the electrons mass fraction    
\begin{align}
 Y_{e^-} &= \dfrac{a^2}{b} \left[1- \tanh^2 \left\{- \dfrac{Le_z a}{2 (m^+/m^-)} (x+{\cal C})
\right\} \right]\\
a &=1- \dfrac{\mu^{1/2}  (m^+/m^-)^{1/2} }{2} E_0\\
b &=4 \dfrac{\mu^{1/2}  (m^+/m^-)^{1/2} }{\varepsilon Le_z},
\end{align}
to be used as a substitute of the boundary condition at $x\to -\infty$ given above.

\label{2}

\section{Numerical method}
The problem defined by Eqs. (\ref{eq:theta_eq})- (\ref{eq:electron_eq}), with the corresponding
boundary conditions, was solved numerically to compute the eigenvalue $\mu$ and the profiles of
temperature and species in a non-uniform grid spanning from $x_{min}=-800$ to $x_{max}=200$ with a
maximum clustering of points around the flame location $x=0$.
The spatial derivatives were discretized using second order, three-
point central differences in a grid formed by $N = 5000$ points, what gives a
minimum spacing $dx \simeq 0.02$ at $x=0$. A 50 \% increase in the number of points was used in some
cases to test the grid independence of the numerical solution. \\
The eigenvalue $\mu$ and the profiles of temperature and species were computed using an iterative
method based on a Gauss–Seidel procedure with over-relaxation that takes advantage of the
invariance of the equations to a translation in the coordinate $x$. Using this property,  a random
value of temperature $\theta^*$ is forced at the grid point $x^*$ such that
$\theta(x^*)=\theta^*$, what gives an additional condition that allows the calculation of the
eigenvalue $\mu^{k}$ at the iteration $k$ from Eq. (\ref{eq:theta_eq}). To avoid the divergence of
the method, we used a relaxation parameter $\alpha$ so that the value of the eigenvalue used at the
next iteration $k+1$ is given by $\mu^{k+1}=\alpha \mu^{k} + (1-\alpha) \mu^{k-1}$. Typical values
of above mentioned parameters are  $\alpha=0.5$ and $\theta^*=0.65$. A comprehensive description of
the numerical procedure outlined above can be found in \cite{Kurdyumov2013}.

\section{Results}
\subsection{Structure}
We begin by examining the basic structure of flames modeled in this formulation as shown in \autoref{fig:profiles_sketch}. The solutions for the neutral species $Y_F$, $Y_Z$ and $\omega_I$ closely follow those found in prior work studying reactions I and II \cite{Dold2007}. $\theta$ initially follows the neutral solution, but develops more slowly in the later stages of the flame. This follows naturally considering that, with small values of ${\cal A}$ and $\Delta \beta \geq 0$, reactions I and II are dominant until $Y_F$ becomes small and temperature overcomes the higher activation energy in reaction III.\\
Once reaction III begins to proceed in earnest, heat release becomes highly dependent upon the presence of both $Y_{Z^+}$ and $Y_{e^-}$. Upon the application of a positive $E_0$ we see significant shifts in the profiles of both ion species. Electron concentration before and within the flame region is elevated, permitting earlier electron-proton recombination. The resulting accelerated heat evolution is subtle but present in $\theta$ and, due to exponential temperature dependence, raises the peak values of $\omega_I$ and $\omega_{III}$. So, we find that by controlling the delayed heat release of the chemiionization path it is possible to realize significant changes in total reaction rate, which is to say, flame speed as is found in \autoref{fig:AA_vs_E}. The opposite effect is observed when a negative electric field is applied. Recombination is delayed, temperature rises more slowly, reaction rates are decreased and flames speed decreases.\\
Observation of ion species only in the flame region is not sufficient to fully account for the processes at work. Particularly, a wider field of view is necessary to understand the local rise and fall of $Y_{e^-}$. The inset of \autoref{fig:profiles_sketch} $b)$ plots the self-induced electric field, from which the net charge may be deduced, over a large domain. In the case where $E_0 \geq 0$, we see that $\p E/ \p x \leq 0$ for $x \leq 0$ while $\p E/ \p x >0$ (weakly) for $x > 0$ (weakly), indicating that electrons are shifted strongly into both the flame sheet and pre-flame regions. This accumulation increases the availability of electrons within the flame sheet to participate in reaction IV and release heat $q_{IV}$ closer to the unreacted fuel. Where $E_0 \leq 0$, electrons are driven far into the post-flame region. This effectively robs the flame sheet of a fraction of $Q$ by separating the components, $Y_{Z^+}$ and $Y_{e^-}$ necessary for reaction IV.\\
Fundamentally, their high mobility causes electrons to be a deficient or limiting component for flame propagation. The influence of $E_0$ and the reason for its direction dependence come from a capacity to oppose or enhance the loss of electrons by advection and diffusion. Electrons driven forward by positive $E_0$ are simply not lost. They accumulate ahead of the flame but are not destroyed and remain available to react within the flame sheet. Electrons driven behind the flame, conversely, quickly become so far removed from the flame sheet that the heat they release cannot contribute meaningfully to propagation. \\
Note that flame speed will not increase without limit as $E_0$ increases.  As seen in Figures \ref{fig:AA_vs_E}-\ref{fig:epsilon_vs_E}, for each set of parameters (${\cal A}$, ${\cal B}$, etc.) there is a critical value of $E_0$ at which flame speed is maximal and further increases in $E_0$ reduce flame speed.\\
\subsection{Parameters}
In the interest of generality, the dimensionless parameters ${\cal A}$, ${\cal B}$, $\varepsilon$ and $Q$ have been kept somewhat arbitrary, but their values do modify the  sensitivity of flames to externally applied electric fields.\\
First, we consider ${\cal A}$ which controls the rate at which $Y_Z$ produces both $Y_{Z^+}$ and $Y_{e^-}$.  Increasing the value of ${\cal A}$ increases both the intensity of the enhancement effect and the critical value of $E_0$.  The first point is unsurprising given that ${\cal A}$ increases the production of species directly influenced by $E_0$.  The second stems from increased self-induced electric fields made possible by higher total concentrations of both ion species.  These self-induced fields naturally oppose the charge separation effect that limits flame speed enhancement.\\
The efficiency with which the ion species recombine is governed by ${\cal B}$.  In the extreme that ${\cal B} \rightarrow \infty$ the recombination would occur instantly, with the release of $q_{IV}$ being limited by reaction III. \autoref{fig:BB_vs_E} shows that as this ${\cal B}$ is increased, ion transport diminishes and  higher electric fields are necessary to achieve similar increases in flame speed. The maximum value of $(\mu(0)/\mu(E))^{1/2}$ increases very weakly with {\cal B}. High rates of recombination can only shift the peak of $\omega_{IV}$ as far forward as that of $\omega_{III}$, which limits the potential for propagation enhancement.\\
As consistent with laminar flame theory and evidenced by prior work on laminar flames \cite{Dold2007, Kurdyumov2013}, increases in the heat release parameter $Q$ can dramatically increase flame speed.  We see in \autoref{fig:Q_vs_E} that this does not change the initial value of  $\p (\mu(0)/\mu(E))^{1/2}/\p E$, but does increase the critical value of $E_0$ and, therefore, the maximum value of $(\mu(0)/\mu(E))^{1/2}$ .  We have discussed the  role of self-induced electrostatic fields in opposing detrimental charge separation, but the constant forward travel of the flame front also serves minimize the relative motion of ions when $E_0>0$.  More simply, one may consider this as a balance between flame speed, $S_L$, and the electromotive component of drift velocity in \autoref{eq:drift}.\\
Lastly, observe in \autoref{fig:epsilon_vs_E} the impact of $\varepsilon$.  Recalling, of course, from \autoref{eq:E} that intensity of the auto-induced field will vary inversely with $\varepsilon$.  $\varepsilon = \infty$ corresponds to ions which are acted upon solely by $E_0$ and diffusion.  As such, flame speed exhibits its strong sensitivity to $E_0$.\\
For small values, of $\varepsilon$, a different phenomenon develops.  When $\varepsilon  \approx 1$, any charge separation is vigorously opposed and transport of $Y_{Z^+}$ by the imposed field can shift the heat release by reaction IV.  Hence we see a reversal in the response of flame speed. A positive $E_0>0$ tends to drive positive ions away from the flame front, delaying heat release.  Similarly, weakly negative values $E_0$ will actually marginally increase flame speeds.\\

\section{Conclusions}
The effect of an external electric field on a freely propagating, planar
and adiabatic premixed flame is investigated for an idealized chemical mechanism that includes
a two-step, chain-branching model and two additional reactions to account for the production and
consumption of charged species. \\
The basic structure of the flames modeled in this paper follows the
structure described by previous studies \cite{Dold2002} in the low temperature
region of the flame. Once $Y_F \ll 1$, the temperature-dependent chemiionization step becomes
dominant and the heat released is then controlled by the concentration of protons and electrons, as
indicated by reaction IV.\\
The application of an external electric field $E_0$ changes the
distribution of protons and electrons around the thin chain-branching layer what, in turn, modifies
the rate at which the heat is released and induces changes in the flame speed. Concretely, the
application of a positive electric field $E_0$ promotes the accumulation of electrons in the cold
region of the flame and increases its availability to react with the protons through reaction IV
once they temperature is sufficiently high to overcome the activation energy of reaction III. The
opposite effect is observed when $E_0<0$.\\
The sensitivity of the flame velocity regarding several of the
non-dimensional parameters of the problem have been tested. Specifically, we focused on the effect of the
frequency factors of reaction III and IV, ${\cal A}$ and ${\cal B}$ respectively, the heat
released parameter $Q$ and the non-dimensional permittivity ${\bar \varepsilon}$. For all the four
parameters, the calculations revealed a decrease of the flame speed for $E_0<0$, due to the reduction
of the electrons concentration before the flame. On the other hand, an increase of the
flame speed is observed for $E_{0,max}>E_0>0$, where $E_{0,max}$ is the maximum
electric field at which $(\mu(0)/\mu(E))^{1/2}>1$. For values of $E_0>E_{0,max}$, the effect of
the electric field on the flame speed is reverted and $(\mu(0)/\mu(E))^{1/2}<1$. The reason for
this is the effective diffusion of electrons towards the cold size of the flame
induced by the electric field, what reduces the rate of the dissociative recombination reaction and
delays the rise of temperature
behind the flame. \\
Maximum flame speed increments of around 15 \% have been found for a specific combination of the non-dimensional parameters of the problem. This number is modest when compared with the most extreme experimental data found in the literature \cite{Murphy2013}, but agrees with the saturation effect reported in \cite{Jaggers1971} for lean flames.  The model presented is not a tool for quantitative prediction, but a tool for developing an improved qualitative understanding of the phenomenon.  This can serve as a basis for progressively building less abstract models by including additional features, such as thermal expansion or a more complex chemical kinetics.

\section*{Acknowledgements}
\addcontentsline{toc}{section}{Acknowledgements}
\label{sec:acknowledgements}
This collaborative research was supported by the Spanish MCINN under Project \#ENE2012-33213 and by King Abdullah University of Science and Technology (KAUST), Cooperative Agreement \# 025478 entitled, “Electromagnetically Enhanced Combustion: Electric Flames”

\label{sec:references}
\bibliographystyle{model1a-num-names}
\bibliography{Scale}
\addcontentsline{toc}{section}{References}







\clearpage

\begin{table*}[htb]
\centering
\small
\begin{tabular}{r l  m{20pt} l l l}
\hline
 i$_{\rm \sc SD}$ & Reaction                &            &           \\ \hline
 \\
  I  & F+Z           $\rl$ 2 Z               &            & $\Omega_I= A_I \dfrac{\rho^2}{W_F
W_Z} Y'_Z Y'_F \exp\left(-E_I/R T \right)$\\
\\
  II  & Z+M           $\rl$ P + M +q$_{II}$     &            & $\Omega_{II}= A_{II}
\dfrac{\rho^2}{W_Z
\bar{W}} Y'_Z$\\
\\
 III  & Z+M           $\rl$ Z$^{+}$ + e$^{-}$ + M     &            & $\Omega_{III}= A_{III}
\dfrac{\rho^2}{W_Z
\bar{W}} Y'_Z \exp\left(-E_{III}/R T \right)$ \\
\\
IV  & Z$^{+}$+ e$^{-}$       $\rl$ P+ q$_{IV}$     &            & $\Omega_{IV}=
A_{IV} \dfrac{\rho^2}{W_{Z^{+}} W_{e^{-}}} Y'_{Z^{+}} Y'_e$  
\end{tabular}
\caption{Chain-branching, chemionizaition and dissociative recombination reactions used in this formulation}
\label{table:reactions}
\end{table*}

\begin{figure*}[!ht]
\begin{center}
\includegraphics[width=.45\textwidth,angle=-90]{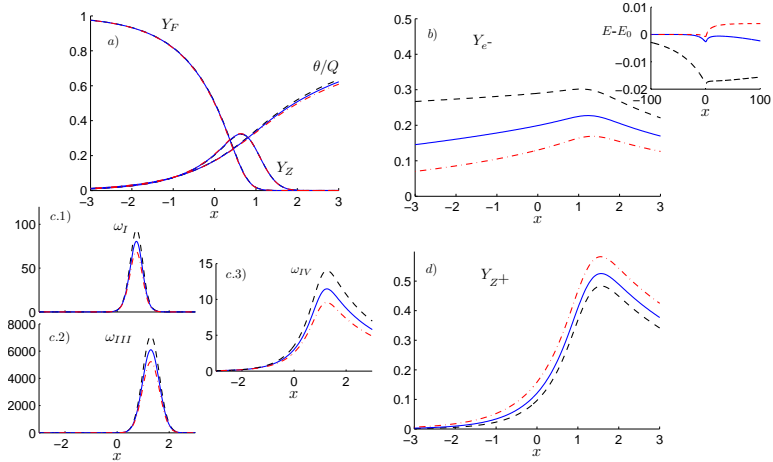}
\caption{a) Profiles of fuel $Y_F$, radical $Y_Z$ and temperature $\theta/Q$, b) and d)
Electron and proton mass fraction profiles, c.1) chain-branching reaction rate $\omega_I$, c.2)
proton-electron production rate $\omega_{III}$ and c.3) proton-electron recombination reaction rate.
$\omega_{IV}$ for $E_0=0$ (blue, solid line), $E_0=0.65$ (black, dashed line) and
$E=-0.65$ (red dot-dashed lines) with $\beta=10, \Delta \beta=1, {\cal A}=0.1, {\cal B}= 100$,
$\varepsilon=100$, $m^+/m^-=100$ and $Q=5$. The inset of figure 1b) represents the auto-induced electric field created by the charge displacement near the flame.}
\label{fig:profiles_sketch}
\end{center}
\end{figure*}

\begin{figure}[!ht]
 \includegraphics[width=.45\textwidth,angle=-90]{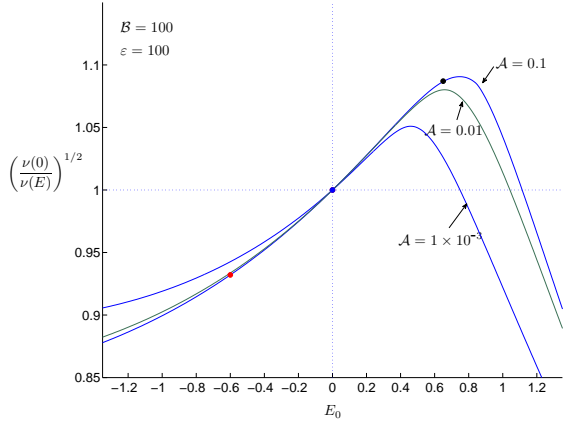}
 \caption{Influence of ${\cal A}$ on Flame speed vs. electric field $E_0$ for  $\beta=10, \Delta \beta=1,
{\varepsilon}=100, {\cal B}= 100$, $Q=5$, $m^+/m^-=100$. Points for ${\cal A}=0.1$ correspond to the cases in Figure \ref{fig:profiles_sketch}} 
\label{fig:AA_vs_E}
\end{figure}

\begin{figure}[!ht]
 \includegraphics[width=.45\textwidth,angle=-90]{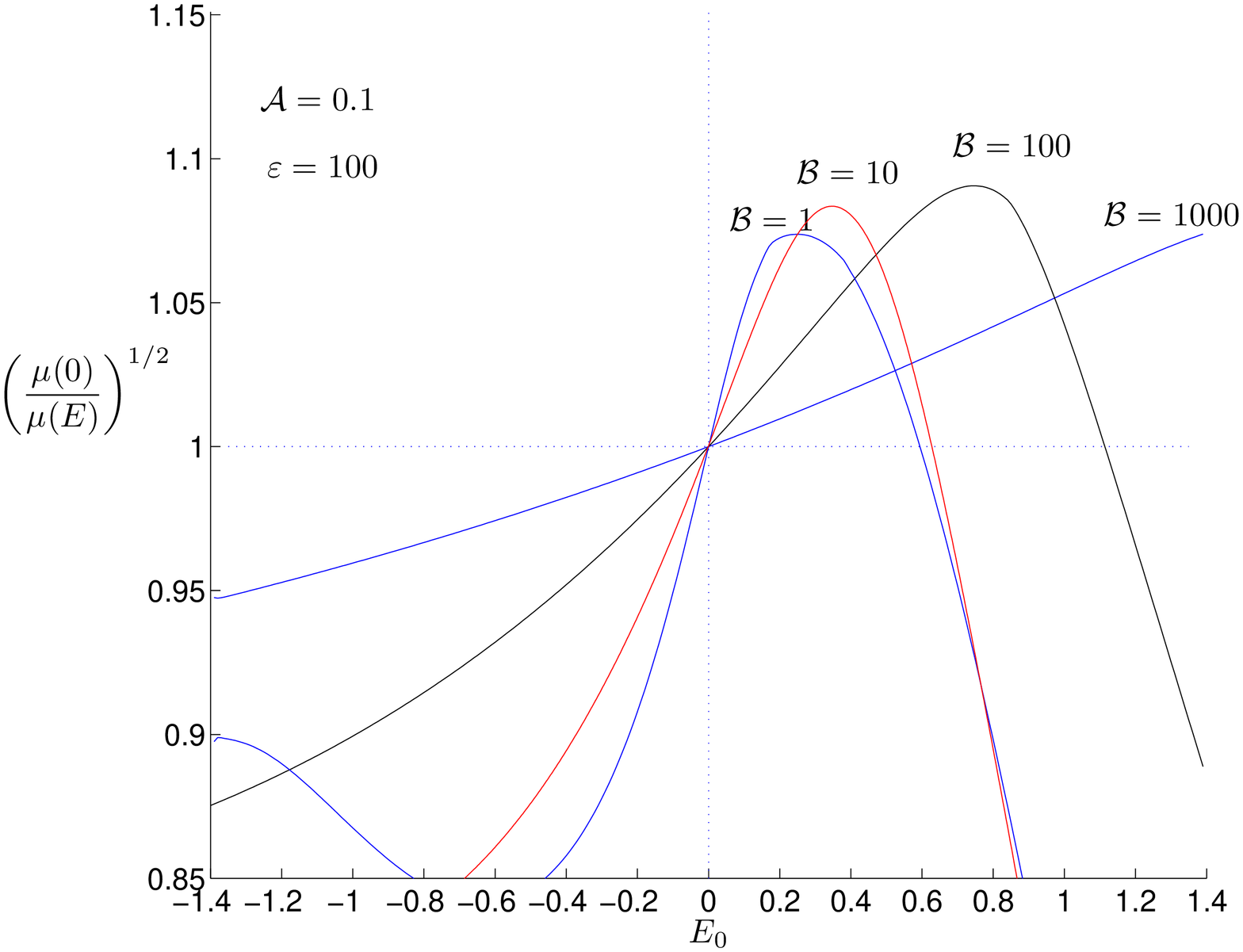}
 \caption{Influence of ${\cal B}$ on Flame speed vs. electric field $E_0$ for  $\beta=10, \Delta \beta=1,
{\varepsilon}=100, {\cal A}= 0.1$, $Q=5$, $m^+/m^-=100$. } 
 \label{fig:BB_vs_E}
\end{figure}

\begin{figure}[!ht]
 \includegraphics[width=.45\textwidth,angle=-90]{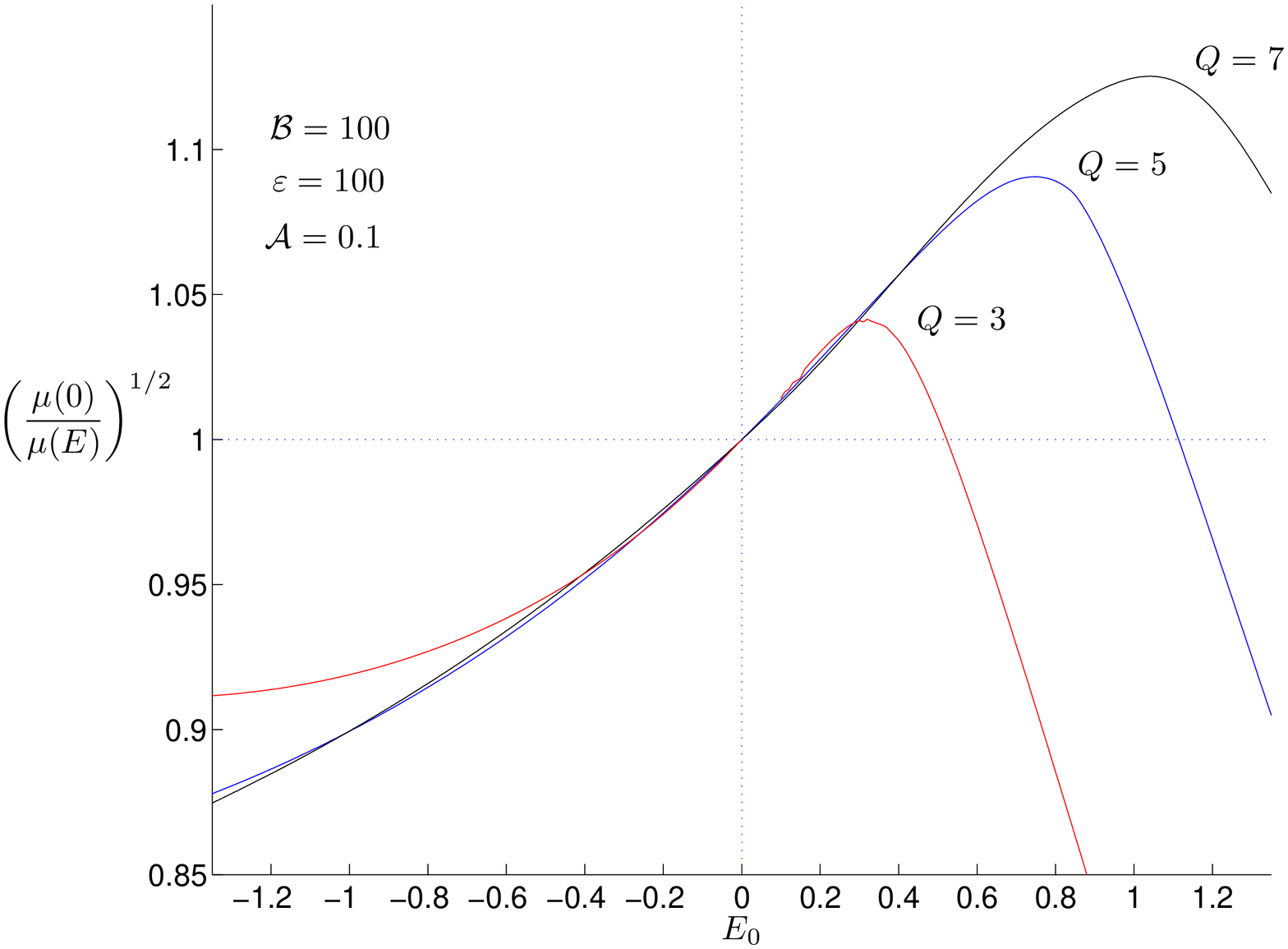}
 \caption{Influence of $Q$ on Flame speed vs. electric field $E_0$ for  $\beta=10, \Delta \beta=1,
{\varepsilon}=100, {\cal A}= 0.1,{\cal B}= 100$, $m^+/m^-=100$. } 
 \label{fig:Q_vs_E}
\end{figure}

\begin{figure}[!ht]
 \includegraphics[width=.45\textwidth,angle=-90]{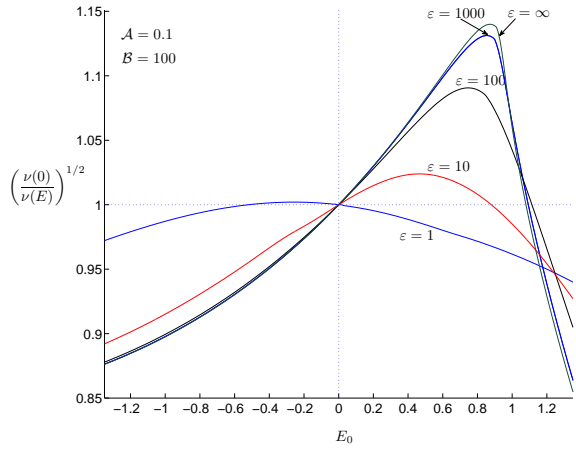}
 \caption{Influence of $\varepsilon$ on Flame speed vs. electric field $E_0$ for  ${\cal B}= 100, {\cal A}= 0.1$, $Q=5$, $m^+/m^-=100$. } 
 \label{fig:epsilon_vs_E}
\end{figure}

\clearpage

\pagebreak[4]
{\Large Figure Captions}

\autoref{fig:profiles_sketch} a) Profiles of fuel $Y_F$, radical $Y_Z$ and temperature $\theta/Q$, b) and d)
Electron and proton mass fraction profiles, c.1) chain-branching reaction rate $\omega_I$, c.2)
proton-electron production rate $\omega_{III}$ and c.3) proton-electron recombination reaction rate
$\omega_{IV}$ for $E_0=0$ (blue, solid line), $E_0=0.65$ (black, dashed line) and
$E=-0.65$ (red dot-dashed lines) with $\beta=10, \Delta \beta=1, {\cal A}=0.1, {\cal B}= 100$,
$\varepsilon=100$, $m^+/m^-=100$ and $Q=5$. The inset of figure 1b) represents the auto-induced electric field created by the charge displacement near the flame.

\autoref{fig:AA_vs_E} Influence of ${\cal A}$ on Flame speed vs. electric field $E_0$ for  $\beta=10, \Delta \beta=1,{\varepsilon}=100, {\cal B}= 100$, $Q=5$, $m^+/m^-=100$. Points for ${\cal A}=0.1$ correspond to the cases in Figure \ref{fig:profiles_sketch}

\autoref{fig:BB_vs_E} Influence of ${\cal B}$ on Flame speed vs. electric field $E_0$ for  $\beta=10, \Delta \beta=1,{\varepsilon}=100, {\cal A}= 0.1$, $Q=5$, $m^+/m^-=100$.

\autoref{fig:Q_vs_E} Influence of $Q$ on Flame speed vs. electric field $E_0$ for  $\beta=10, \Delta \beta=1, {\varepsilon}=100, {\cal A}= 0.1,{\cal B}= 100$, $m^+/m^-=100$. 

\autoref{fig:epsilon_vs_E} Influence of $\varepsilon$ on Flame speed vs. electric field $E_0$ for  ${\cal B}= 100, {\cal A}= 0.1$, $Q=5$, $m^+/m^-=100$.

\end{document}